\documentclass[8.5pt,twoside,twocolumn]{article}
\usepackage[super,sort&compress,comma]{natbib} 
\usepackage[version=3]{mhchem}
\usepackage{times,mathptmx}
\usepackage{sectsty}
\usepackage{balance} 

\usepackage{graphicx} 
\usepackage{lastpage}
\usepackage[format=plain,justification=raggedright,singlelinecheck=false,font=small,labelfont=bf,labelsep=space]{caption} 
\usepackage{fancyhdr}
\begin{document}

\thispagestyle{plain}
\fancypagestyle{plain}{
\fancyhead[L]{\includegraphics[height=8pt]{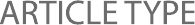}}
\fancyhead[C]{\hspace{-1cm}\includegraphics[height=20pt]{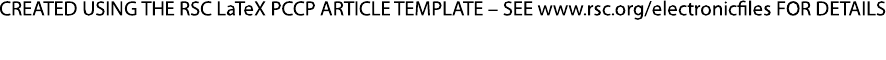}}
\fancyhead[R]{\includegraphics[height=10pt]{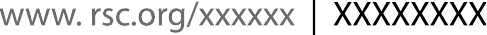}\vspace{-0.2cm}}
\renewcommand{\headrulewidth}{1pt}}
\renewcommand{\thefootnote}{\fnsymbol{footnote}}
\renewcommand\footnoterule{\vspace*{1pt}%
\hrule width 3.4in height 0.4pt \vspace*{5pt}} 
\setcounter{secnumdepth}{5}

\makeatletter 
\def\subsubsection{\@startsection{subsubsection}{3}{10pt}{-1.25ex plus -1ex minus -.1ex}{0ex plus 0ex}{\normalsize\bf}} 
\def\paragraph{\@startsection{paragraph}{4}{10pt}{-1.25ex plus -1ex minus -.1ex}{0ex plus 0ex}{\normalsize\textit}} 
\renewcommand\@biblabel[1]{#1}            
\renewcommand\@makefntext[1]%
{\noindent\makebox[0pt][r]{\@thefnmark\,}#1}
\makeatother 
\renewcommand{\figurename}{\small{Fig.}~}
\sectionfont{\large}
\subsectionfont{\normalsize} 

\fancyfoot{}
\fancyfoot[LO,RE]{\vspace{-7pt}\includegraphics[height=9pt]{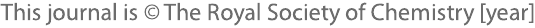}}
\fancyfoot[CO]{\vspace{-7.2pt}\hspace{12.2cm}\includegraphics{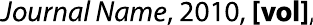}}
\fancyfoot[CE]{\vspace{-7.5pt}\hspace{-13.5cm}\includegraphics{RF}}
\fancyfoot[RO]{\footnotesize{\sffamily{1--\pageref{LastPage} ~\textbar  \hspace{2pt}\thepage}}}
\fancyfoot[LE]{\footnotesize{\sffamily{\thepage~\textbar\hspace{3.45cm} 1--\pageref{LastPage}}}}
\fancyhead{}
\renewcommand{\headrulewidth}{1pt} 
\renewcommand{\footrulewidth}{1pt}
\setlength{\arrayrulewidth}{1pt}
\setlength{\columnsep}{6.5mm}
\setlength\bibsep{1pt}

\twocolumn[
  \begin{@twocolumnfalse}
    \noindent\LARGE{\textbf{Structural Transitions and Hysteresis in Clump- and Stripe-Forming Systems Under Dynamic Compression}} 
\vspace{0.6cm}

\noindent\large{\textbf{Danielle McDermott,\textit{$^{1,2}$}  
Cynthia J. Olson Reichhardt,$^{\ast}$\textit{$^{1}$} 
and Charles Reichhardt\textit{$^{1}$}}}\vspace{0.5cm}

\noindent\textit{\small{\textbf{Received Xth XXXXXXXXXX 20XX, Accepted Xth XXXXXXXXX 20XX\newline
First published on the web Xth XXXXXXXXXX 20XX}}}

\noindent \textbf{\small{DOI: 10.1039/b000000x}}
\vspace{0.6cm}

\noindent \normalsize{Using numerical simulations, 
we study the dynamical  
evolution of 
particles interacting via 
competing long-range repulsion and short-range attraction                
in two dimensions.
The particles are compressed  
using a time-dependent 
quasi-one dimensional trough potential 
that 
controls the local density, 
causing 
the system to undergo a series of 
structural phase transitions 
from
a low density clump lattice 
to stripes, voids, and a high density uniform state.
The compression proceeds 
via slow elastic motion 
that is interrupted with 
avalanche-like bursts of activity 
as the system collapses 
to progressively higher densities 
via plastic rearrangements.  
The plastic events 
vary in magnitude from small rearrangements of particles, 
including the formation of quadrupole-like defects,  
to large-scale vorticity and 
structural phase transitions.
In the dense uniform phase, 
the system 
compresses through row reduction transitions
mediated by
a disorder-order process.
We characterize the rearrangement events 
by measuring
changes in the potential energy, 
the fraction of sixfold coordinated particles, 
the local density, and
the velocity distribution.
At high confinements, we find power law scaling of
the velocity distribution 
during 
row reduction transitions.  
We observe hysteresis under a reversal of the compression 
when relatively few plastic rearrangements occur.
The decompressing system
exhibits 
distinct phase morphologies,
and the phase transitions occur at lower compression forces as the system
expands compared to when it is compressed.
}
\vspace{0.5cm}
  \end{@twocolumnfalse}
]

\section{Introduction}

\footnotetext{\textit{
$^1$~Theoretical Division,
    Los Alamos National Laboratory, Los Alamos, New Mexico 87545 USA.
    Fax: 1 505 606 0917; Tel: 1 505 665 1134; E-mail: cjrx@lanl.gov}}
\footnotetext{\textit{
    $^2$~Department of Physics,
    Wabash College, Crawfordsville, Indiana 47933 USA.
    Fax: 1 765 361 6340; Tel: 1 574 361 6305; E-mail: mcdermod@wabash.edu}} 

Superlattices of clump crystals, stripes, and voids
form in a variety of
soft matter systems \cite{1,2,3,4,5,6,7}, 
hard solids \cite{8,9,10,11}
and dense nuclear matter \cite{12,13}.
These systems can often be modeled as a collection of particles 
with pairwise interactions that act over multiple
length scales.
For instance, 
a rich variety of phases 
appear for
a particle interaction potential that 
combines 
long range repulsion, 
which favors a uniform triangular crystal, 
with short range attraction,  
which favors condensed structures \cite{3,4}. 
Clump and stripe morphologies
also appear
for strictly repulsive particle interactions 
provided that multiple length scales are present,
such as in two-step repulsive potentials \cite{14,15}.
More generally, 
systems in which the Fourier transform
of the particle interaction potential contains a negative mode can exhibit
clump or stripe phases \cite{16} that can undergo 
a multiple-step
melting transition in which the ordering is destroyed at one length scale
but preserved at another \cite{5}.
Introduction of a confining potential to 
such pattern forming systems
can generate additional ordering effects
such as the formation of rings, bands 
and other symmetric patterns \cite{17,18,19,20},
while the presence of a periodic substrate can induce
commensurate-incommensurate transitions
\cite{21}. 
Studies of
pattern forming systems
generally focus on the types of patterned phases
that appear, their equilibrium properties,
and 
the nature of transitions 
between phases as a function of temperature, density, and 
interparticle interactions.
Far less is known about 
the nonequilibrium behaviors of these systems 
under external drives, 
shear, 
or dynamic compression. 
In previous work, studies of the dynamical ordering of patterns
driven over random substrates \cite{22,23,24,25,26,27} 
or under an applied shear \cite{28} 
showed that
the system
tends to form stripes that align in the direction of the drive or shear.

In this work
we numerically examine the evolution of a system
of particles with 
competing long range repulsion and short range attraction  
that are dynamically compressed and 
subsequently decompressed by a
quasi-one-dimensional confining potential. 
In the absence of a confining potential, the particles
organize into a clump state.
During compression, 
the system undergoes 
a series of structural transitions 
from clumps to stripes, 
stripes to voids, 
and voids to a dense uniform phase
by means of
large-scale particle rearrangements.
These large events are interspersed with
smaller scale rearrangements
consisting of localized excitations
that are often quadrupolar in form.
The structural rearrangements 
produce changes in 
the number of sixfold coordinated particles,
the effective local density, and 
other geometric and energetic measures.
We find a broad distribution
of particle velocities
with multiple scaling regimes.
In the highly compressed dense uniform phase,
the particle velocity distribution has avalanche-like
power law scaling when large-scale rearrangements
associated with changes in the triangular lattice ordering
occur.
These avalanche-like events,
which we call row reduction transitions, 
occur through an order-disorder compression mechanism
that differs from 
the dynamical behavior found for
purely repulsive particles under compression \cite{38}.
The system exhibits hysteresis
under decompression
when the phase transitions shift to
lower confining forces.
Additionally,
different stripe and void structures appear under decompression since
fewer plastic particle rearrangements
occur compared to the compression process.

Our model of a confining trap could 
be created experimentally
through the optical trapping of soft matter systems \cite{29,30,31},
which permits the dynamical tuning of the strength of the confining potential.
There are examples of Coulomb crystals 
in which particle confinement is changed 
by varying the strength of the confinement  
as a function of time \cite{32,33,34,35}.
Another compression method 
involves placing a  
soft matter assembly between two walls 
that confine the particles but allow fluid to flow through,
such as 
in a recent bubble raft experiment 
on particles with competing interactions 
in which compression was used
to achieve a structural transition from a less rigid to a more rigid amorphous phase 
that was correlated with a change  
in the mean coordination number and displacement field \cite{36}.   
It may also be possible to compress a packing of
magnetic particles using a changing magnetic field \cite{37}.


\section{Simulation and System}
We consider a two-dimensional (2D) system of size $L \times L$
with periodic boundary conditions in the $x$ and $y$-directions containing
$N=256$ particles
whose pairwise interactions include
both repulsive and attractive components.
The particle density is given by $\rho=N/L^2$.
The particle
dynamics are governed by the following overdamped equation
of motion:
\begin{equation}
\eta \frac{d {\bf R}_{i}}{dt} =
-\sum^{N}_{j \neq i} \nabla V(R_{ij}) + {\bf F}^{s}_{i} + {\bf F}^T_i.
\end{equation}
Here $\eta$ is the damping term which we set to unity and
${\bf R}_{i (j)}$ is the location of particle $i (j)$.
The particle-particle interaction potential
has the form $V(R_{ij}) = 1/R_{ij} - B\exp(-\kappa R_{ij})$,
where $R_{ij}=|{\bf R}_i-{\bf R}_j|$.
The Coulomb term $1/R_{ij}$ produces a repulsive interaction at
long range, 
while the exponential term gives
an attraction at shorter range. 
At very short range,
the repulsive Coulomb interaction becomes dominant again.
We place a cutoff on the interactions at $R_{ij} < 0.1$ 
to avoid the divergence of the Coulomb interaction, and use
a Lekner summation method to treat
the long-range Coulomb forces \cite{39}.
The particles are confined by
a single trough potential 
${\bf F}^{s}_i = F_{p}\cos(2\pi x_i/L){\hat {\bf x}}$ 
which exerts an $x$-direction force that pulls the particles
toward a central minimum.
The thermal fluctuations ${\bf F}_i^T$ are present only during annealing
and not during compression/decompression, and represent Langevin kicks with
the properties $\langle F_i^T\rangle=0$ and
$\langle F_i^T(t)F_j^T(t^\prime)\rangle=2\eta k_BT\delta_{ij}\delta(t-t^\prime)$,
where $k_B$ is the Boltzmann constant.
Previous studies \cite{4} of this model in the absence of a substrate
showed that for fixed $B = 2.0$ and $\kappa = 1.0$,
at low densities $\rho\leq 0.27$ the system forms clumps that grow in size with
increasing $\rho$.
A stripe state appears
for $0.27 < \rho \leq 0.46$,
void crystals form
for $0.46 < \rho \leq 0.58$,
and a uniform triangular lattice
appears for $\rho > 0.58$.
Here we fix $B=2.0$ and $\kappa=1.0$.
We take $L = 36.5$ in dimensionless simulation length units so that
in the absence of a confining potential the particle density is
$\rho = 0.19$.

Before we apply compression forces, we anneal
the system from a high temperature 
molten state down to zero temperature in small increments
in order to obtain an initial substrate-free
uniform distribution of clumps. 
Starting with the $T=0$ substrate-free annealed system, we ramp
the substrate strength from $0 \le F_p \le 10$
in quasistatic increments of
$\Delta F_p = 0.001$ every
$\Delta t = 2000$ simulation time steps, which permits the system
to reach equilibrium at each confinement force,
and then lower $F_p$ back to zero at the same rate. 
For comparison, we also perform a static anneal of the system
at selected fixed values of $F_p$ in order to identify 
when the dynamically compressed
particles are trapped in a metastable configuration.

During each confining force increment we measure 
$\rho_{\rm eff}$, 
the density of the particles trapped within the well, given by
\begin{equation}
\rho_{\rm eff} = \frac{N}{L (x_{\rm max}-x_{\rm min})}, 
\end{equation}
where $x_{\rm max}$ ($x_{\rm min}$) is the $x$-position of the rightmost
(leftmost) particle
in the sample.
Since the particles are pointlike, $\rho_{\rm eff}$ can become quite
large in the densely compacted state.
We also measure 
$\sigma_{xx}$, 
an element of the stress tensor:
\begin{equation}
\sigma_{xx} = \frac{1}{L^2} \sum_{i}^{N}\sum_{j<i}{ (\vec{F}_{ij})_x (\vec{R}_{ij})_x} ,
\end{equation}
where $\vec{F_{ij}}$ 
is 
the interparticle force between particles $i$ and $j$
and $\vec{R}_{ij}$ is their relative separation.
Further measures include 
the 
change in the total particle-particle interaction energy 
with compression force, $dE/dF_p$, where 
$E=\sum_{i}^{N}\sum_{j\neq i}^{N}V(R_{ij})$, and 
the local ordering $P_6=N^{-1}\sum_{i=1}^N\delta(z_{i}-6)$, where $z_i$ is the
coordination number of particle $i$ obtained from a Voronoi tessellation.
We
do not include particles along the sample edges
in our measurement of $P_6$ since they may show a local structure
consistent with sixfold ordering
despite not actually having six neighbors.

\begin{figure*}[t]
  \centering
  \includegraphics[width=\textwidth]{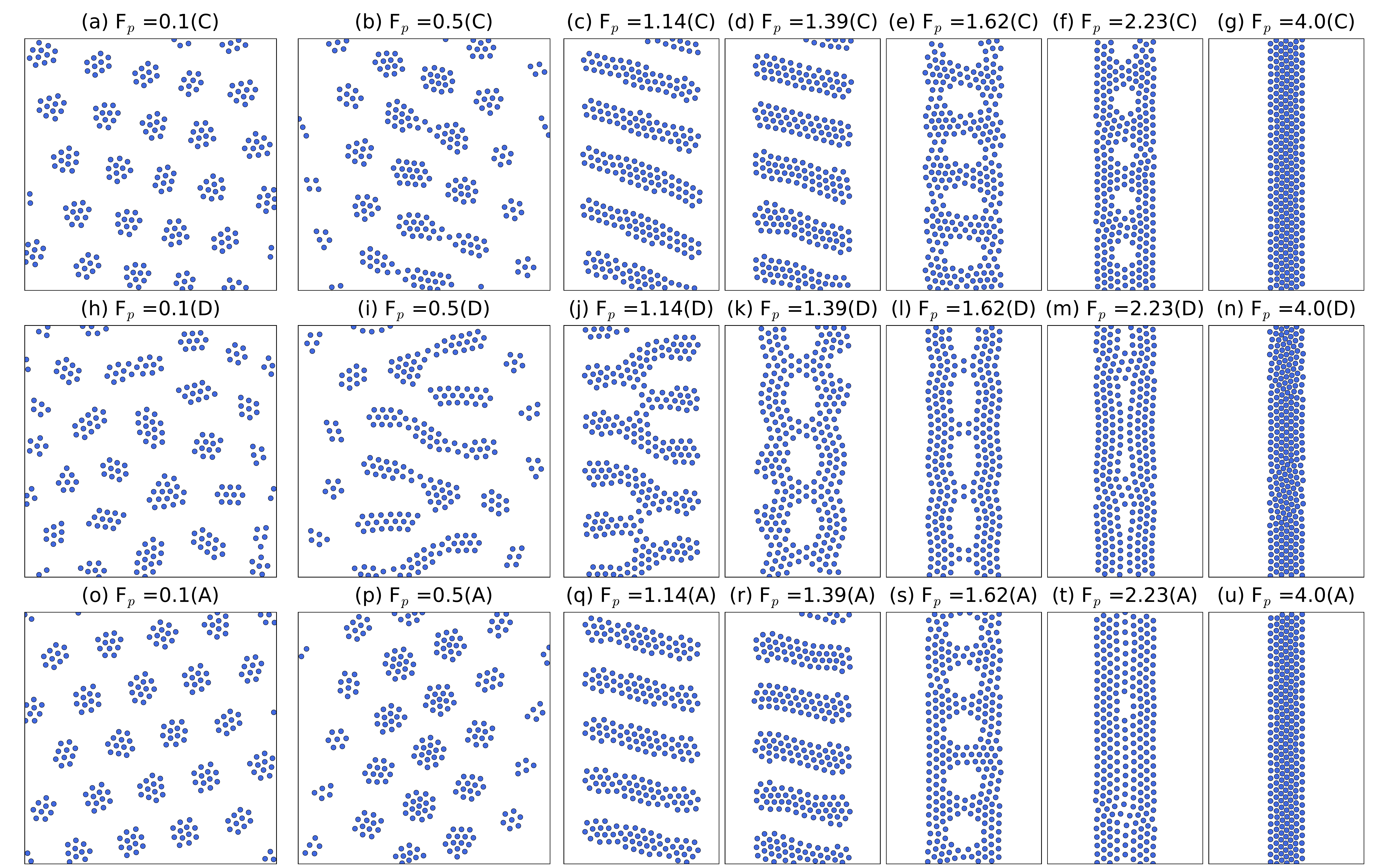}
\caption{ 
(a-u) Particle positions at increasing substrate strength $F_p$ from left to right.  Top row (a-g): compression (C); middle row (h-n): decompression (D); bottom row (o-u): annealed with fixed confining force (A).  The confining forces  $F_p$ are: 0.1 (a,h,o), 0.5 (b,i,p), 1.14 (c,j,q), 1.39 (d,k,r), 1.62 (e,l,s), 2.23 (f,m,t), and 4.0 (g,n,u).  The supplementary materials fully illustrate the dynamics of compression \cite{supp1} and decompression \cite{supp2}.
}
\label{fig:1}
\end{figure*}

\section{Results}
In Fig.~\ref{fig:1} we plot the 
particle positions
at selected confinement forces 
from compression (C, top row) and decompression (D, middle row) sequences as well as 
for 
static annealing (A, bottom row).
The annealed configurations
represent the ground state of the system, 
while the compression and decompression sequences
exhibit hysteresis 
due to metastability caused by kinetic trapping.

Figure~\ref{fig:1}(a-g) illustrates
the density-dependent phase transitions that occur during
compression.
The dynamics
of these transitions are shown fully in supplementary materials \cite{supp1}.
In Fig.~\ref{fig:1}(a), at
$F_p=0.10$, the sample is filled with clumps that move
as stable raftlike objects which interact only
through long range repulsive forces.
Each clump contains approximately ten particles
with local triangular ordering.
At $F_p=0.50$ in 
Fig.~\ref{fig:1}(b),
the clumps interact and rearrange, 
becoming elongated as they converge at the center of the trap 
into proto-stripes.
Figure~\ref{fig:1}(c)
shows that
the stripes 
at $F_p=1.14$
are roughly three particles wide, are slightly thicker near the trap center, and
span the trap diagonally 
as in Ref. ~\cite{21}. 
As in Ref.~\cite{4},
the stripe width is approximately equal to
the diameter of the annealed clumps, 
a length scale that is determined by the range of the attractive interaction.
As $F_p$ continues to increase, 
the stripes 
become narrower in the $x$-direction, 
as shown in Fig.~\ref{fig:1}(d) for $F_p=1.39$.
Here the stripes are four particles in width and exhibit some thickening
at their outer edges.
After the stripes have collapsed, as shown in
Fig.~\ref{fig:1}(e) at $F_p=1.62$,
five evenly-spaced voids form 
in the areas that previously separated
the stripes. 
There are many defects and grain boundaries 
in the local particle ordering
due to the rapid convergence of 
the separated stripes into a 
connected solid.
Upon further compression 
these defects gradually heal as 
the voids shrink and become circular,
as shown in Fig.~\ref{fig:1}(f) at $F_p=2.23$.
Finally, at $F_p=2.4$,  
the voids collapse
and the particles form a disordered dense solid
that is $n=8$ rows wide (not shown).
The particles eventually adopt nearly crystalline ordering
in the dense phase, 
as illustrated in Fig.~\ref{fig:1}(g) at $F_p=4.0$.

\begin{figure*}
  \centering
\includegraphics[width=0.61\textwidth]{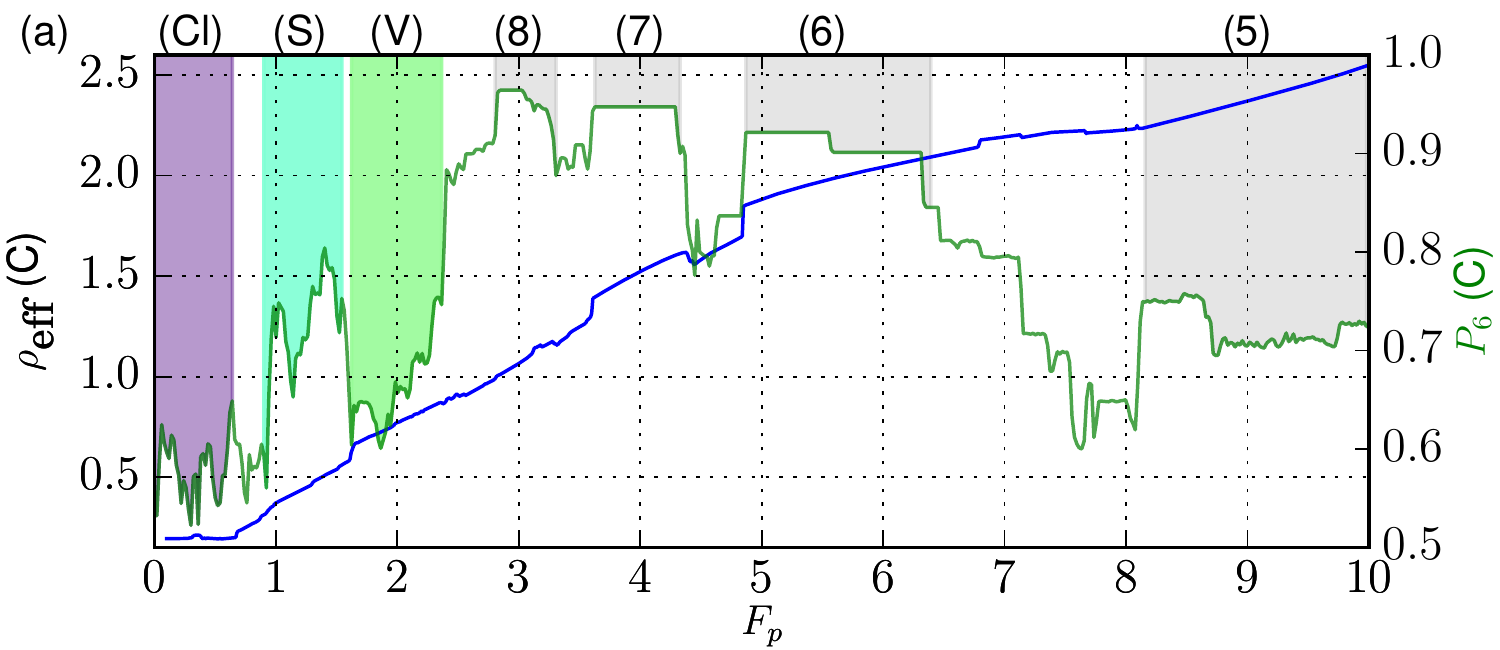}
\includegraphics[width=0.3\textwidth]{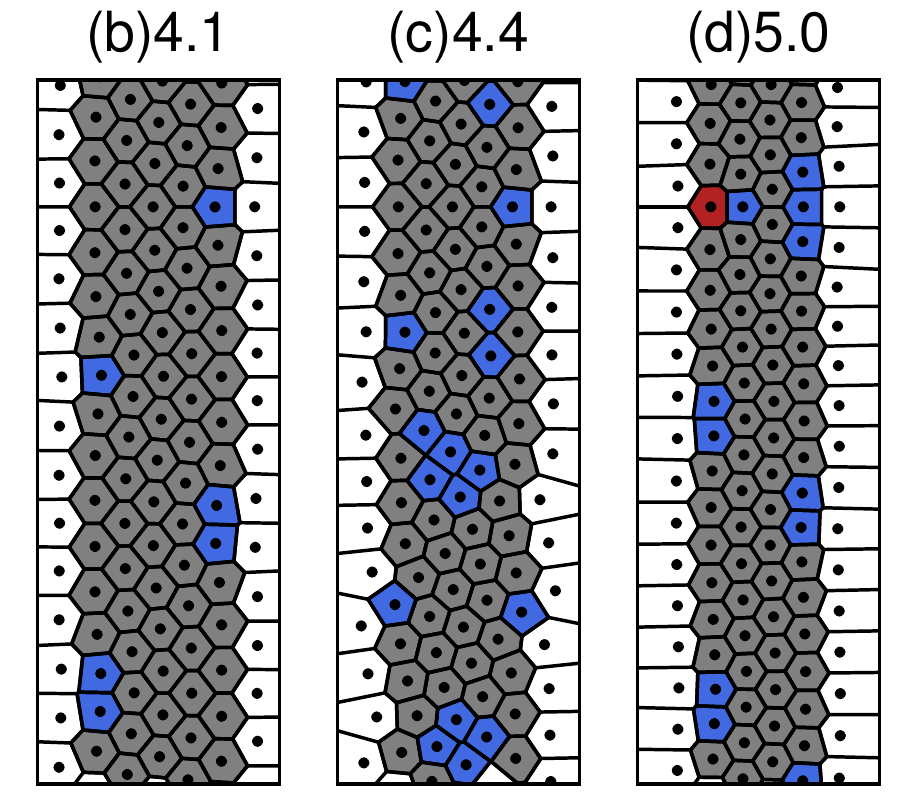}
\caption{(a) $\rho_{\rm eff}$ (blue) and $P_6$ (green) vs $F_p$ illustrating the transitions that occur under compression.  The background shading indicates the regions of clumps (Cl, lavender), stripes (S, blue), voids (V, green), and crystalline order ($n$, gray) with white representing mixed phases or disordered regions.  Each shaded region is labeled on the top axis, where the numbers in the crystalline ordered regime indicate that the packing is $n$ rows wide.
(b-d) Voronoi tessellation images of a portion of the compressed system illustrating the order-disorder cycle that occurs during an $n=7$ to $n=6$ row reduction transition.  Voronoi cells with five sides are blue, six sides are gray, and seven sides are red. (b) $F_p=4.1$. (c) $F_p=4.4$.  (d) $F_p=5.0$.
}
\label{fig:2}
\end{figure*}

The qualitative behavior under decompression
is similar to that observed for compression,
as shown in Fig.~\ref{fig:1}(h-n);
however,
we find hysteresis in the onset of the 
transitions and the detailed morphology of the phases.
The decompression dynamics
are fully illustrated in the supplementary materials 
\cite{supp2}.
In Fig.~\ref{fig:1}(n) at $F_p=4.0$,
multiple triangularly ordered domains appear in the decompressing
$n=7$ row wide dense packing
as part of a
transition in the
number $n$ of rows in the dense packing.
As expansion continues,
voids form by a nucleation process 
in which the dense solid 
unzips, 
creating 
two long gaps at the substrate minimum as
shown in Fig.~\ref{fig:1}(m) 
at $F_p=2.23$.
Although the compression of the void state 
involves localized 
defects and plastic rearrangements,
during decompression
the edge particles move 
smoothly and symmetrically outward.
The symmetric voids at $F_p=1.62$, 
illustrated in Fig.~\ref{fig:1}(l),
transition into the asymmetric arrangement shown in
Fig.~\ref{fig:1}(k) 
at $F_p=1.39$  
when
a small void-like region opens
on each edge of the sample.
The asymmetric voids
unravel into W-shaped labyrinth-like stripes that
span the trap
diagonally in both directions, as
shown 
in Fig.~\ref{fig:1}(j) for $F_p=1.14$,
instead of forming the uniformly spaced
stripes that appear during  compression.
The stripe to clump transition is shifted to lower $F_p$
during decompression,
compared to where it occurs under compression,
and proceeds via the formation of a mixed clump-stripe
state of the type illustrated
in Fig.~\ref{fig:1}(i) for $F_p=0.50$.
The final clump state, shown
in Fig.~\ref{fig:1}(h) at $F_p=0.10$, 
is not uniform 
but has a clear size distribution, with larger clumps located
closer to the substrate minimum. 
The distribution of clump sizes is a result of the
metastable persistence of the stripes
to lower confinement during decompression.

In Fig.~\ref{fig:1}(o-u) we show
the particle configurations obtained by annealing
the particles in a static
potential with fixed $F_p$.
At
$F_p=0.1$ in Fig.~\ref{fig:1}(o),
the annealed system is nearly identical to
the dynamically compressed system of Fig.~\ref{fig:1}(a).
At $F_p=0.5$ in 
Fig.~\ref{fig:1}(p),
the stripelike
order found under compression and decompression is replaced by inhomogeneous
clumps,
with
the largest clumps located closest to the substrate minimum.
In the annealed stripe state at $F_p=1.14$ in
Fig.~\ref{fig:1}(q),
the modulated stripe widths minimize
the repulsive interaction between the particles,
and the stripes are
narrower than those that form under compression.
At $F_p=1.39$, where a metastable
void state appears under decompression,
Fig.~\ref{fig:1}(r)  shows that
the annealed sample 
forms stripes that are slightly thickened at the outer edges,
similar to the compressed state at the same $F_p$ in Fig.~\ref{fig:1}(d).
In contrast to
the elongated voids that
form under compression or decompression,
the annealed voids,
shown in Fig.~\ref{fig:1}(s) 
at $F_p=1.62$,
are nearly circular,
while the particles
surrounding the voids form a partially disordered triangular lattice.
Figure~\ref{fig:1}(t) illustrates
the last vestiges of the annealed void state 
at $F_p=2.23$ 
where a disordered triangular lattice appears that has a reduced
density near the substrate minimum.  The well-formed voids
shown in Fig.~\ref{fig:1}(f) in
the compressed sample at the same $F_p$ are thus metastable.
At $F_p=4.0$ in Fig.~\ref{fig:1}(u),
the ordering in the dense annealed state is indistinguishable from that
in the compressed packing in Fig.~\ref{fig:1}(g).

In Fig.~\ref{fig:2}(a)  
we illustrate the dynamical evolution 
of $P_6$ and $\rho_{\rm eff}$
under compression.
We observe clump (Cl), stripe (S), and
void (V) ordering, as well as a dense $n$ row triangular lattice
($n$) with $n$ ranging from 8 to 5.
At low $F_p$,
$P_6$ fluctuates rapidly and shows
large jumps at the morphological transitions into and out of ordered states,
while at high $F_p$,
there are a series of plateau regions 
with $P_6>0.9$
interrupted by sharp drops in $P_6$ 
when row
reductions
occur.
differ from
the morphological phase transitions
that occur at low confinement.
When the particles are compressing via
slow elastic motion,
$\rho_{\rm eff}$ 
changes smoothly,
whereas a
sudden decrease of the width of the packing
produces a
sudden jump in  $\rho_{\rm eff}$. 
Often such jumps coincide with
jumps in $P_6$ 
since ordered crystalline domains with high $P_6$
appear when a relatively disordered 
diagonal patch of the type illustrated
in Fig.~\ref{fig:1}(n)  
reorders under the applied compression,
so that the packing simultaneously becomes narrower and better ordered.

For low compression forces $0.0<F_p<0.26$,
$\rho_{\rm eff}$ in Fig.~\ref{fig:2}(a) remains flat
since the particles still span the entire width of the
system, and
the clumps of particles move as stable rafts 
until the first clump-clump collision occurs.
There is a large upward jump in $P_6$ and a slight increase
in $\rho_{\rm eff}$ at $F_p\approx 0.66$ at the end of
the pure clump phase
where 
the first stripe forms.
Under further compression,
$\rho_{\rm eff}$ steadily
increases
and shows
occasional jumps produced by structural transitions.
For $0.9<F_p<1.6$,
the particles rapidly migrate toward the
minimum of the substrate potential,
and 
the increased local density
causes a transition
to an ordered stripe phase  (S).
The average value of $P_6$ is noticeably
higher in the stripe state than in the clump or void states,
and $P_6$ drops
sharply at $F_p=1.6$ at the stripe to void transition,
where
a system-wide avalanche occurs and $\rho_{\rm eff}$ shows a slight jump.
For $1.6<F_p<2.4$, 
compression proceeds via
the gradual shrinking of the voids.
Several jumps in $P_6$ occur 
as particles fill the outer edges of the void regions,
while $\rho_{\rm eff}$ gradually increases.
The closure of the voids 
is marked by a large jump in $P_6$ 
and small fluctuations in $\rho_{\rm eff}$
when the system forms a disordered $n=8$
solid band
that has a slightly reduced particle density at the substrate minimum.

As the compression proceeds in the densely packed phase, 
the particles shift toward the substrate minimum and
develop sixfold ordering,
as indicated by the slow rise in $P_6$ 
in Fig.~\ref{fig:2}(a) over the range $2.4 < F_p < 2.9$.  
The sharp jump in $P_6$ at $F_p=2.9$ 
occurs when the packing
first becomes nearly crystalline. 
As the number of rows $n$ decreases step by step during
compression, $P_6$ alternates from
$P_6>0.9$
when the sample is in an $n$ row ordered state
to
$P_6<0.9$ when the sample disorders
during the $n$ to $n-1$ transition.
In a study of the compression of
purely repulsive particles in Ref.~\cite{38},
the primary mechanism for row reduction transitions 
was a release from a highly anisotropic to a relatively isotropic 
triangular ordering via edge defect formation.
In contrast, for the pattern forming system,
the row reductions occur gradually
due to the
short range interparticle attractive force, and are
initiated by
a necking effect consisting of a localized density increase, 
as illustrated in the Voronoi tessellations of Fig.~\ref{fig:2}(b-d) and in
the supplementary videos ~\cite{supp1}.
For $3.6<F_p<4.2$, there is a plateau
in Fig.~\ref{fig:2}(a) where $P_6 \approx 0.96$,
indicating that the system is nearly crystalline,
as illustrated at $F_p=4.1$ in Fig.~\ref{fig:2}(b).
As the compression continues,
additional row reductions occur, until
at $F_p=10.0$, 
the particles are in a highly
compressed state with $n=5$,
at which point we begin the
decompression process.

\begin{figure}[t]
\includegraphics[width=3.5in]{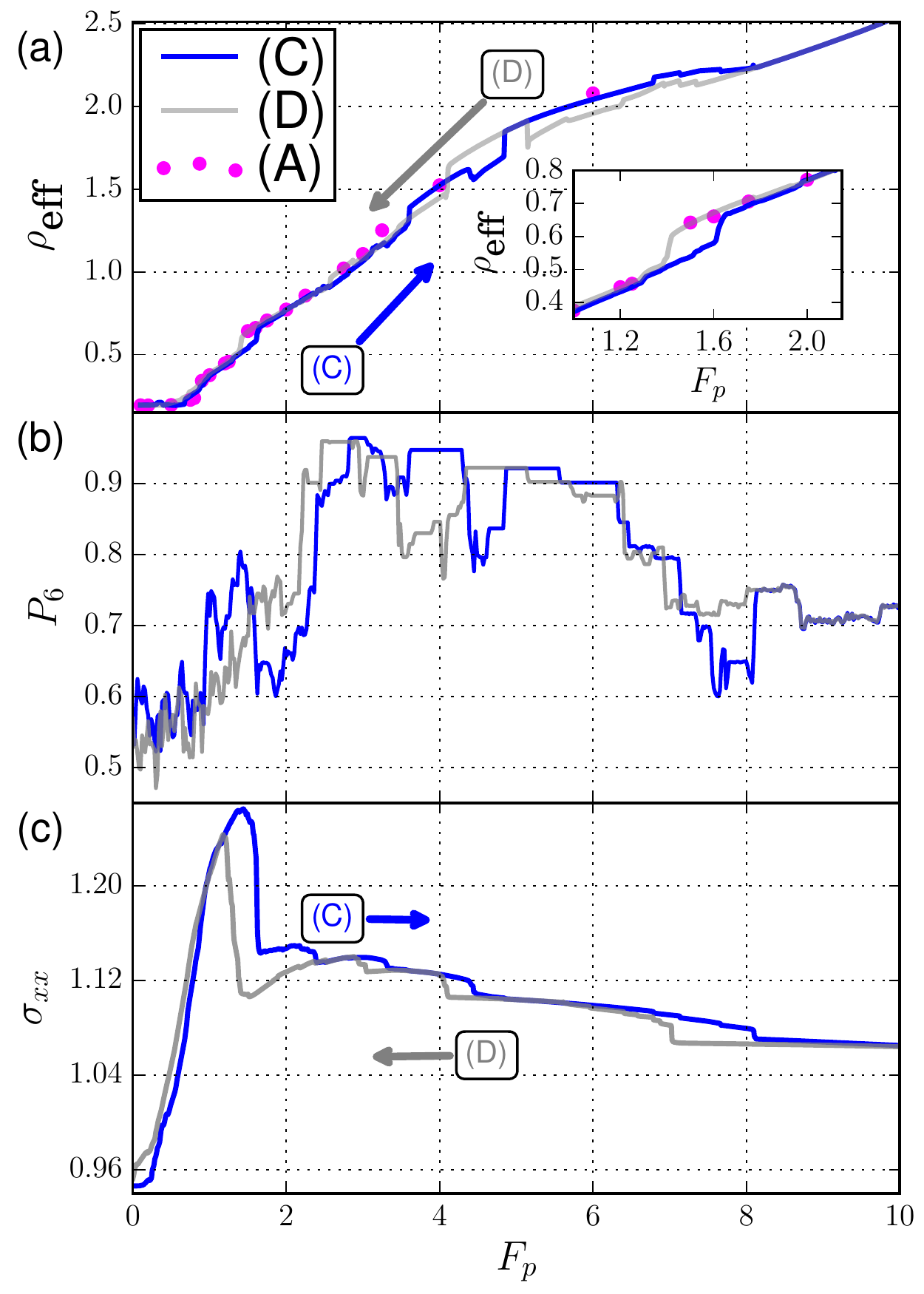}
\caption{(a) $\rho_{\rm eff}$ vs $F_p$, (b) $P_6$ vs $F_p$, and (c) $\sigma_{xx}$ vs $F_p$ for the system under compression (blue), decompression (gray), and static annealing (magenta).  The compression curves in panels (a) and (b) also appear in Fig.~\ref{fig:2}.  In the inset of panel (a) we highlight the hysteresis in $\rho_{\rm eff}$ at the stripe-void transition.
}
\label{fig:3}
\end{figure}

The hysteretic signatures in $\rho_{\rm eff}$, $P_6$, and $\sigma_{xx}$ versus $F_p$
for compression and decompression are plotted in
Fig.~\ref{fig:3}. 
During decompression,
the structural transitions
are systematically shifted to lower $F_p$
and fewer plastic rearrangements occur.
In Fig.~\ref{fig:3}(a) 
we plot 
$\rho_{\rm eff}$ versus $F_p$ 
for compression,
decompression,
and static annealing.
Under annealing, 
the particles pack more tightly into the substrate minimum
to give the highest values of $\rho_{\rm eff}$.
At high $F_p$,
the decompression of the
dense solid
proceeds via a slow elastic expansion in which 
the particles tend to
maintain the same nearest neighbors
and
the system forms 
inhomogeneous domains
similar to those observed under compression
that 
mediate the disorder-order transitions from
$n$ to $n+1$ rows.
These transitions produce
dips and jumps in $P_6$ in Fig.~\ref{fig:3}(b).
Since only elastic motion occurs for $F_p > 8$,
$\rho_{\rm eff}$, $P_6$, and $\sigma_{xx}$ 
are not hysteretic in this region.

In Figure~\ref{fig:3}(b), the largest single jump in $P_6$
under compression
occurs 
when the voids close at $F_p=2.4$.
There is a complimentary
sharp drop in $P_6$ at $F_p=2$
under decompression
when the voids open.
From $F_p=2.0$ to $F_p=1.4$
in the symmetric void phase under decompression, $\sigma_{xx}$ decreases
with decreasing $F_p$ and $P_6$ fluctuates around a value of $P_6 \approx 0.75$.
Below $F_p=1.4$ for decompression, 
$P_6$ drops
steadily and contains 
no
signatures associated with the formation of
labyrinth-like stripes or the clump state.
When asymmetric voids appear,
for $1.4 > F_p > 1$,
$\sigma_{xx}$ increases with decreasing $F_p$, 
and there is a step in $\rho_{\rm eff}$.
A peak in $\sigma_{xx}$ occurs at the void to stripe transition
at $F_p=1.6$ under compression,
while $\sigma_{xx}$ increases with decreasing $F_p$ under decompression
in both the asymmetric void and the labyrinth phases
before merging with the compression curve
at $F_p=1.4$ 
when the connections between adjacent labyrinths break.
The inset of Fig.~\ref{fig:3}(a) highlights the
hysteresis in $\rho_{\rm eff}$
at the void-stripe transition
where
the jump in $\rho_{\rm eff}$ associated with void collapse/formation
occurs at $F_p=1.6$ under compression and at $F_p=2.0$ under
decompression.  There is an additional drop in $\rho_{\rm eff}$
for the decompressing system at $F_p=1.4$, when the voids change
from symmetric to asymmetric.
The slight hysteresis in $\rho_{\rm eff}$ 
above $F_p=1.6$ is the result of slightly different void morphologies under
compression and decompression, as illustrated in Fig.~\ref{fig:1}.
The effective system width
cannot
capture
the behavior at low confinement forces, so
$\rho_{\rm eff}$
shows no hysteresis for $F_p<1.4$. 

At lower $F_p$,
Fig.~\ref{fig:3}(b) shows that
sharp jumps in $P_6$ appear only during compression-induced
transitions, while
$P_6$ decreases steadily with decreasing $F_p$ under decompression.
In addition,
$\sigma_{xx}$ is systematically higher under expansion than under contraction,
as shown in Fig.~\ref{fig:3}(c).
During decompression, 
the domains tend to undergo elastic
motions
that preserve the nearest neighbor structure of each particle,
so $\sigma_{xx}$ decreases smoothly under decompression
as $F_p$ is lowered and the stripes break apart into clumps,
in a manner similar to the smooth increase of $\sigma_{xx}$
with increasing $F_p$ in the clump
phase under compression.

\begin{figure}[t]
\includegraphics[width=3.5in]{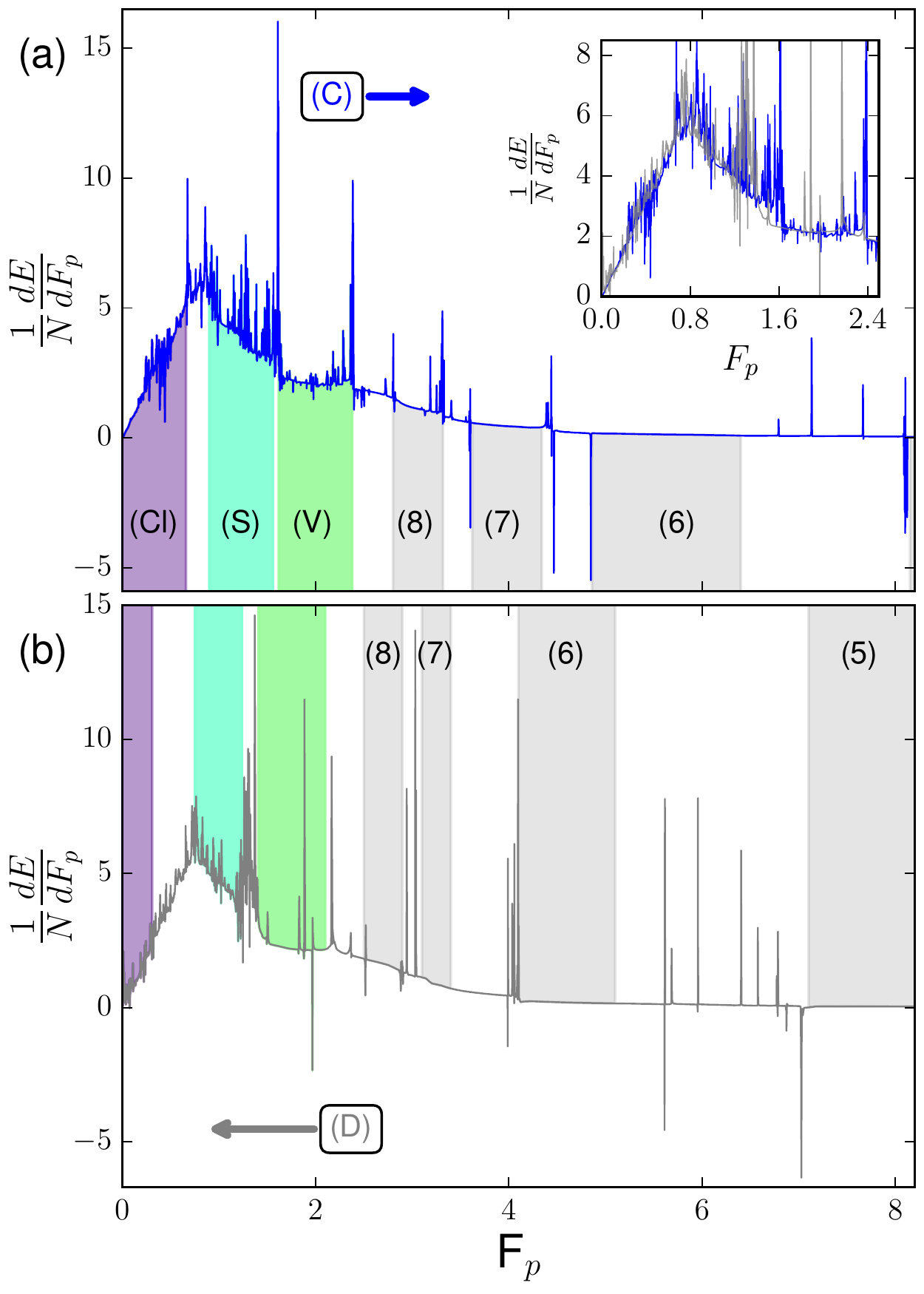}
\caption{(a,b) Change in system energy $dE/dF_p$ vs $F_p$ under compression (a) and decompression (b).  Inset of panel (a): both curves plotted together.  The bands of color denote: clumps (Cl, lavender), stripes (S, light blue), voids (V, green), and dense crystal with different numbers of rows of particles ($n$, gray).
}
\label{fig:4}
\end{figure}

In Fig.~\ref{fig:4}(a,b)
we plot 
the change in the total interparticle
interaction energy $dE/dF_p$ versus $F_p$ 
under compression and decompression, and show both curves
together in the inset of Fig.~\ref{fig:4}(a).Shading indicates the values of $F_p$ for which the sample is in
the clump (Cl), stripe (S), void (V), or crystalline state with $n$ rows of particles ($n$).
White areas indicate mixed regions,
such as the mixed clump-stripe phase 
that appears under decompression
for $0.3<F_p<0.9$.
Spikes in $dE/dF_p$ indicate plastic rearrangements.
For low confinement forces, 
$dE/dF_p$ fluctuates rapidly 
about its average value as the particles
repeatedly rearrange plastically in response
to the changing substrate.
These noise bursts are more intense during compression than during
decompression,
indicating the less frequent occurrence of
plastic rearrangements.
At high $F_p$, the particles are too dense to rearrange easily,
so $dE/dF_p$ becomes smooth
except for rare large avalanche events.

\begin{figure*}
\makebox[\textwidth][c]{\includegraphics[width=0.9\textwidth]{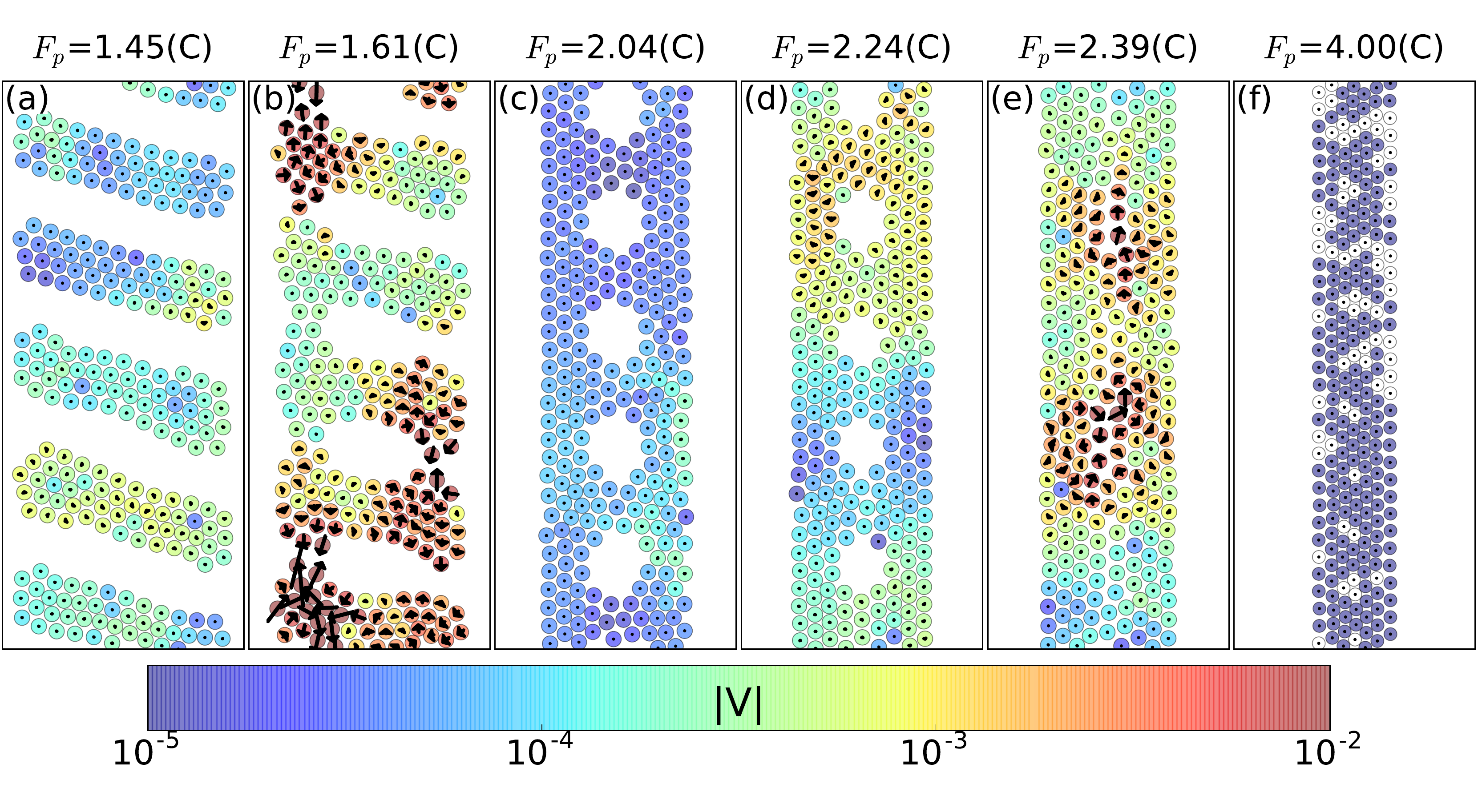}}
\caption{(a-f) Positions and velocities of the compressed system during the formation and collapse of the void state.  The velocity vector (arrow) indicates the instantaneous direction of motion and the particle color indicates speed.  White particles are completely motionless, blue particles are nearly motionless, and red particles move the most rapidly. (a) $F_p=1.45$. (b) $F_p=1.61$. (c) $F_p=2.04$. (d) $F_p=2.24$. (e) $F_p=2.39$. (f) $F_p=4.0$.
}
\label{fig:5} 
\end{figure*}

In Fig.~\ref{fig:4}(a), 
$dE/dF_p$ increases smoothly with $F_p$ at very low $F_p$
due to the increase in the repulsive interparticle interaction as
the clumps glide toward the substrate minimum
before the first clump collision occurs.
The negative spikes in $dE/dF_p$ that occur
in the clump phase result when 
clumps at the edge of the particle assembly break apart, 
while positive spikes in $dE/dF_p$ are produced by
the collision of clumps near the substrate minimum.
At the clump to stripe transition at $F_p=0.9$,
there is a large positive spike in $dE/dF_p$ along with
a slope change
from positive to negative.  
The average value of $dE/dF_p$ 
decreases with increasing $F_p$ for $F_p>0.9$
since the stripe structures thicken while the interparticle
distances change very little.
As a result, 
$E$ decreases as 
the particles gain neighbors
that contribute to the attractive interparticle interaction.
Many positive spikes in $dE/dF_p$ occur
in the stripe regime
due to frequent rearrangements of the particles around the stripe edges.
The stripe-void transition is marked by a large 
positive spike in $dE/dF_p$ at $F_p=1.6$ 
produced when particles rush to fill the inter-stripe gaps,
which suddenly increases
the interparticle repulsive force
in a system-wide avalanche. 
In the void phase, 
small positive spikes in $dE/dF_p$
appear when a few particles move inward in order to
gradually shrink the size of the voids.
The end of the void phase is associated with
a large spike in  $dE/dF_p$ 
at $F_p=2.4$ 
caused by particles flooding into the remaining void space.
There are fewer spikes in $dE/dF_p$ in the dense
solid states
since the deformation of the system 
is dominated by elastic motion 
except for occasional plastic avalanches associated with
row reduction transitions.   
Under compression 
the majority of spikes are positive, 
indicating an increase in the system energy 
as the particles are forced into stronger confinement. 
In contrast, large negative spikes occur during decompression 
that coincide with positive jumps in $P_6$,
associated either with an energy release by the destruction of
strain-induced defects, or with a dynamical reordering from
a highly anisotropic to an isotropic 
arrangement of particles.
At the highest confinement forces, $F_p>8$, 
$dE/dF_p$ is featureless
since no rearrangements occur for either compression or decompression.
The noise in $dE/dF_p$ in the clump and stripe phases 
is reduced during expansion 
due to the lack of plastic rearrangements
compared to compression.
There is no large noise spike at the stripe-clump transition
since these phases coexist
under expansion until $F_p \approx 0.3$. 

In Fig.~\ref{fig:5} we show 
the positions and 
average velocity of individual particles 
in the system 
under compression at different $F_p$ values
spanning the stripe to void and void to dense solid states.
The typical mechanism for stripe compression is illustrated in
Fig.~\ref{fig:5}(a), at $F_p=1.45$.
Each stripe gradually becomes 
thicker as particles rearrange through a vortexlike
motion
at the stripe ends, as shown on the left end of the bottommost
complete stripe.
Such vortexlike motion rarely occurs during decompression.
Figure~\ref{fig:5}(b) shows the collapse of 
the stripe state
into a void state at $F_p=1.61$.
At this point, the stripes have reached their maximum thickness,
which is dictated by the length scale of the short range attraction,
so the stripe ends begin to buckle outward, allowing particles to
bridge the inter-stripe gaps, 
as indicated by the red particles with large velocity vectors.
We show a quiescent void state at $F_p=2.04$
in Fig.~\ref{fig:5}(c), 
where the particles form a disordered triangular lattice
surrounding five nearly circular voids.
At high confinement,
the system spends more time in a quiescent state,
and shows less noise in $dE/dF_p$, 
with no large spikes appearing between $1.6<F_p<2.4$ in Fig.~\ref{fig:4}(a). 

In Fig.~\ref{fig:5}(d) at $F_p=2.24$, 
the voids have been compressed to a smaller size,
and the upper void serves as the center of a vortex of
particle motion,
which is a common rearrangement mechanism in 
the compressing void system. 
Similar vortexlike motions do not occur in the expanding void state.
In Fig.~\ref{fig:5}(e), 
at $F_p=2.39$, 
the voids have just closed, 
and the system forms 
a disordered dense solid
through the plastic motion of many particles.
As the compression proceeds, 
the particle motion becomes
increasingly localized and occurs only when
the system has built up enough strain to trigger a
rearrangement avalanche that reduces the energy of the system.
In Fig.~\ref{fig:5}(f) 
we show
a nearly crystalline state
at $F_p=4.0$. 
The outer rows of particles have smectic-like ordering
and do not form a close packed lattice with the 
inner rows of particles due to the curvature of the substrate.

\begin{figure}[t]
\includegraphics[width=3.5in]{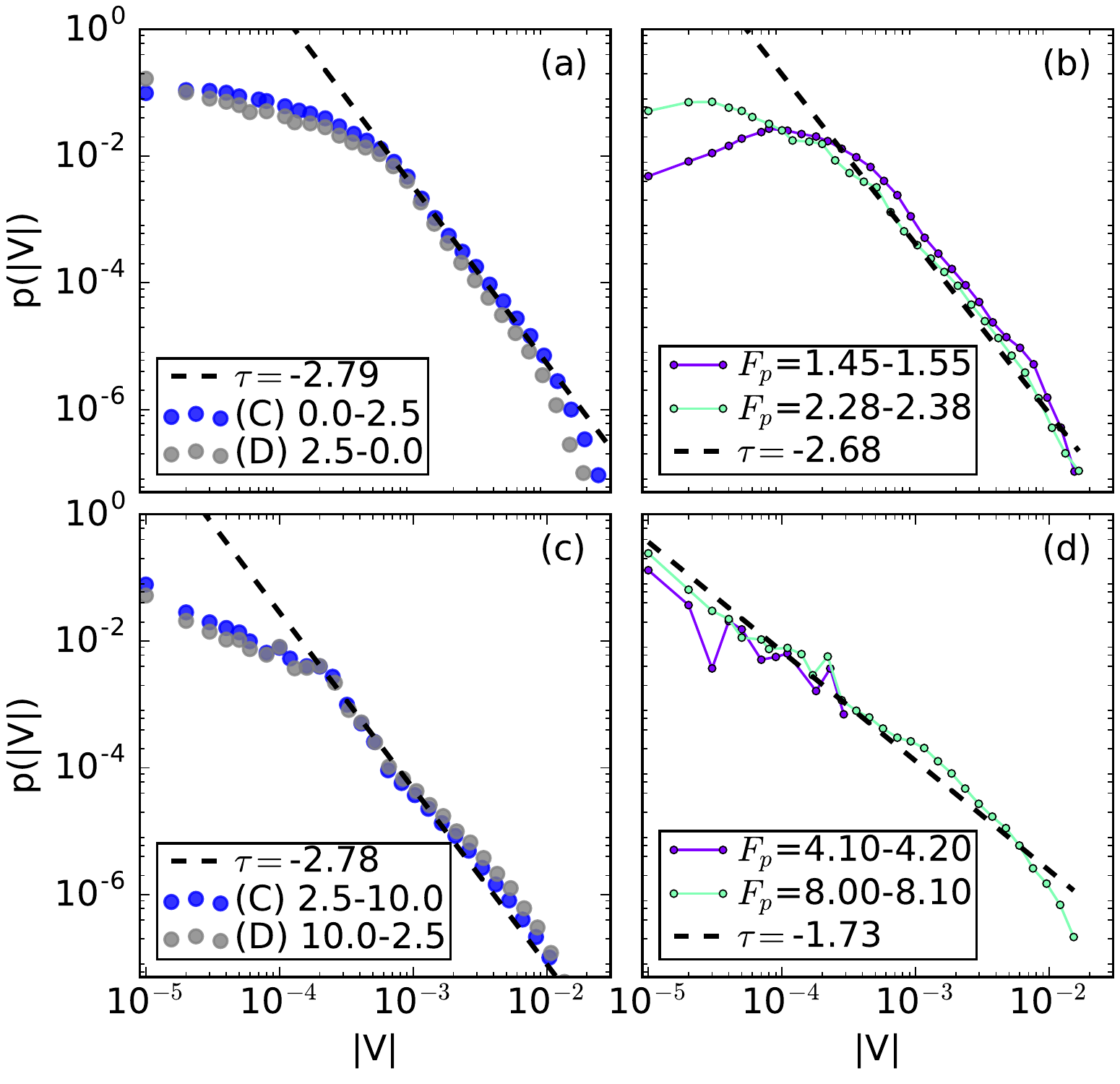}
\caption{(a-d) The velocity distributions $P(|V|)$ vs $|V|$ along with approximate power law fits (dashed) used to estimate the scaling.  (a) Low density phases $0.0 < F_p < 2.5$ under compression (blue) and decompression (gray).  A fit to the intermediate values of the compression data gives an exponent $\tau=-2.79$.  (b) The stripe state with $1.45<F_p<1.55$ (violet) and the void  state with $2.28<F_p<2.38$ (green).  A fit to the tail of both distributions gives $\tau=-2.68$.  (c) The high density phase $2.5 < F_p < 10.0$ under compression (blue) and decompression (gray).  A fit to the tail of the compressed system gives $\tau=-2.78$.  (d) The dense triangular phase under compression in a quiescent period from $4.1 < F_p < 4.2$ (violet) and during a row reduction event $8.0 < F_p < 8.1$ (green).  A fit to all of the data gives $\tau=-1.73$.
}
\label{fig:6}
\end{figure}

In Fig.~\ref{fig:6} we plot the 
velocity distribution
$P(|V|)$ versus $|V|$, 
where $|V|=\sqrt{V_x^2+V_y^2}$, 
on a log-log scale. 
The distribution is broad, 
and
exhibits distinct scaling regimes
produced by different particle behaviors. 
We plot the 
low and high $F_p$ regimes separately
since strong avalanche-like behavior 
is only observed during the row reduction 
transitions of the dense solid state.  
Since the compression is quasistatic, 
the particles are often motionless, 
particularly at high $F_p$.
Velocities in the range
$10^{-5}>|V|>5 \times 10^{-4}$ appear as dark blue
particles in the images of Fig.~\ref{fig:5}, and
undergo motion that is so small as to be negligible.

In Fig.~\ref{fig:6}(a) 
we plot $P(|V|)$ 
in the range $F_p<2.5$ 
for both compression and decompression.
Both curves
have a flat region for $|V|<10^{-3}$, 
an intermediate region $10^{-3}<|V|<10^{-2}$ 
that may exhibit power law scaling, 
and a tail showing a suppression 
of high velocities.
We fit the intermediate region of the 
compressed system 
to a power law, $P(|V|) \propto |V|^{\tau}$, and obtain
an exponent of $\tau =-2.79$.  
For the expanding system, $P(|V|)$ is systematically
shifted to lower values compared to the compressed system,
and the tail of the distribution falls off more rapidly.
This is expected  since the system 
does not exhibit sharp phase transitions at low $F_p$ under decompression,
and therefore the particles move at lower velocities.
It is also consistent with the observation 
that the spikes in $dE/dF_p$ 
at low $F_p$ are less frequent and of smaller magnitude
under expansion than under compression.
In Fig.~\ref{fig:6}(b) 
we plot 
$P(|V|)$ for the most active portions of the compressed system.
In the stripe phase, $1.45 \le F_p \le 1.55$, the stripes are
undergoing the thickening behavior illustrated
in Fig.~\ref{fig:5}(a) in which 
slow elastic compression is interrupted
by small plastic rearrangements.
Here, $P(|V|)$ is relatively low
in the regime of negligible motion $|V|<10^{-4}$, 
has linear scaling over the range $10^{-3}<|V|<10^{-2}$, 
and drops rapidly to zero for
$|V|<10^{-2}$.  This shape resembles that found 
for $F_p<2.5$.
In contrast, in the void regime with $2.28<F_p<2.38$, 
$P(|V|)$ is higher both at low and at high $|V|$. 
A fit to the entire tail of the distribution 
gives $\tau = -2.68$.

In Fig.~\ref{fig:6}(c) 
we plot $P(|V|)$ in the
dense solid state from $2.5<F_p<10$
for both compression and expansion.
There is a distinct regime $|V|>2 \times 10^{-4}$
over which a power law fit gives
$\tau = -2.78$.
Compression and expansion produce nearly the same
$P(|V|)$ curves, although there is slightly more weight in the
tail of the distribution under decompression.
This is consistent with the behavior of $dE/dF_p$ in Fig.~\ref{fig:4},
where the expanding system has
a larger number of higher magnitude spikes
for $F_p>2.5$.
In Fig.~\ref{fig:6}(d) we plot $P(|V|)$
for the compressed system in the range $4.1<F_p<4.2$,
where the particles are nearly crystalline and show no significant
motion.
There are no velocities higher than $|V|=3 \times 10^{-4}$,
while the behavior at low $|V|$
is the same as that observed over the entire range of
$F_p>2.5$.
Figure~\ref{fig:6}(d) 
also shows $P(|V|)$ for a row reduction transition over the range
$8.0<F_p<8.1$,
where a power law fit to the entire distribution gives $\tau = -1.73$.  
We find similar behavior for the other row reduction 
transitions that occur at $F_p \approx 4.3$, 4.9, and $7.1$.
We analyzed a larger system with $L=50.0$ containing
$N=475$ particles, 
and found similar velocity distributions.
The dynamics
of these transitions are shown fully in supplementary materials \cite{supp3}.

\begin{figure}
\includegraphics[width=\columnwidth]{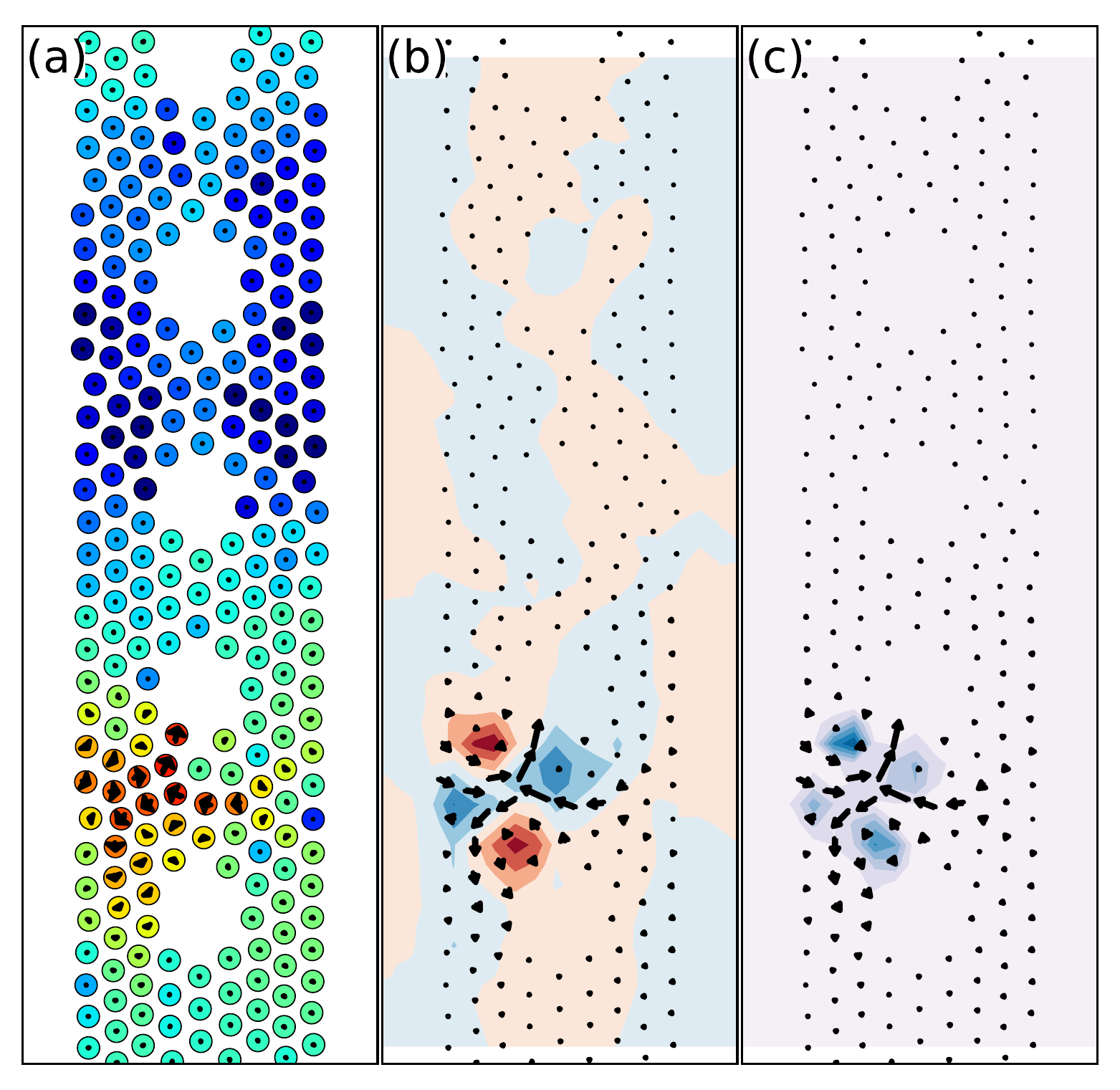}\\
\includegraphics[width=\columnwidth]{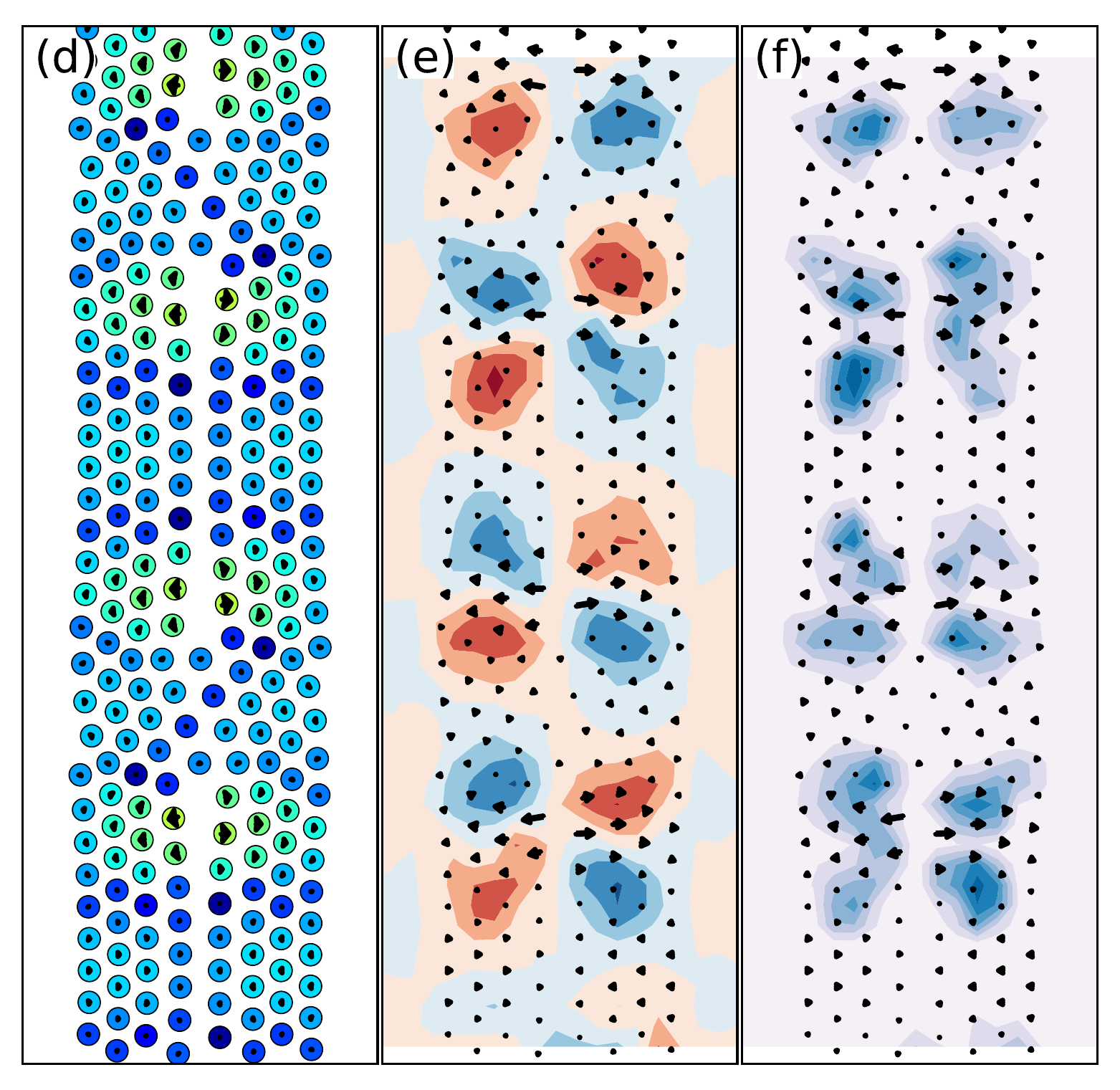}
\caption{(a-c) Quadrupole defect formation under compression at $F_p=2.28$.  (a) Particle location and velocity, with the same coloring as in Fig.~\ref{fig:5}.  (b) The curl of the velocity field with particle velocity vectors.  Blue indicates clockwise curl and red indicates  counterclockwise curl.  (c) The enstropy of the velocity field with particle velocity vectors, where dark regions indicate high enstropy. 
(d-f) The system under decompression at $F_p=2.24$, with the same coloring as above.  (d) Particle location and velocity.  (e) The curl of the velocity field with particle velocity vectors.  (f) The enstropy of the velocity field with particle velocity vectors using the same colors previously described, but with different scaling.
}
\label{fig:7}
\end{figure}

\begin{figure}
\includegraphics[width=0.5\textwidth]{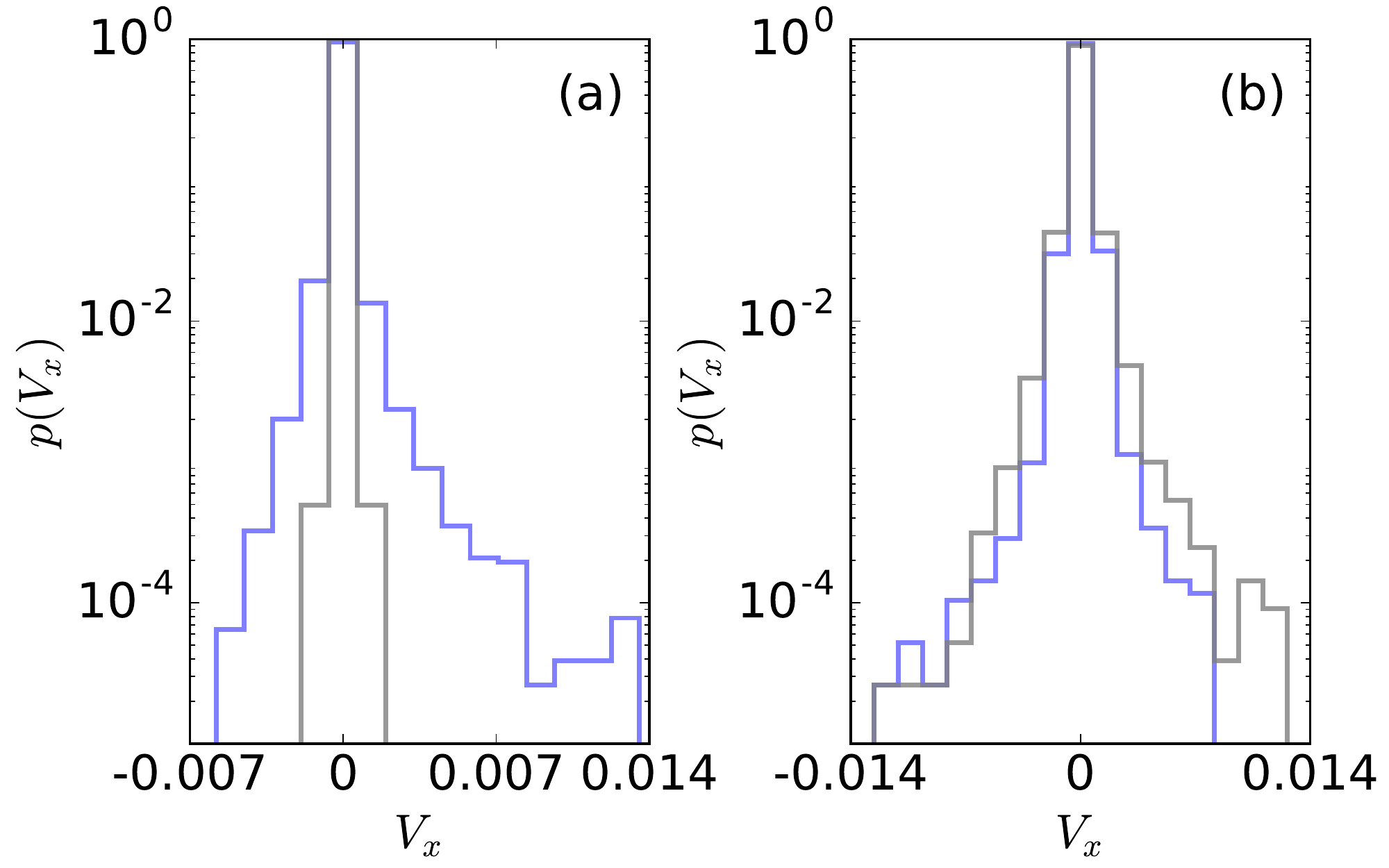}
\caption{(a-b) The velocity distribution $P(V_x)$ of the quadrupole defects on a log-linear scale accumulated over a range $\Delta F_p=0.03$ of substrate strength.  (a) $F_p = 2.28$ (blue) under compression, as illustrated in Fig.~\ref{fig:7}(a-c), and $F_p = 2.24$ (gray) under decompression, as illustrated in Fig.~\ref{fig:7}(d-f).  (b)  $F_p = 1.1$ (blue) under compression, as illustrated in Fig.~\ref{fig:9}(b), and $F_p = 1.0$ (gray) under decompression, as shown in Fig.~\ref{fig:9}(d).
}
\label{fig:8}
\end{figure}

\begin{figure}
\includegraphics[width=\columnwidth]{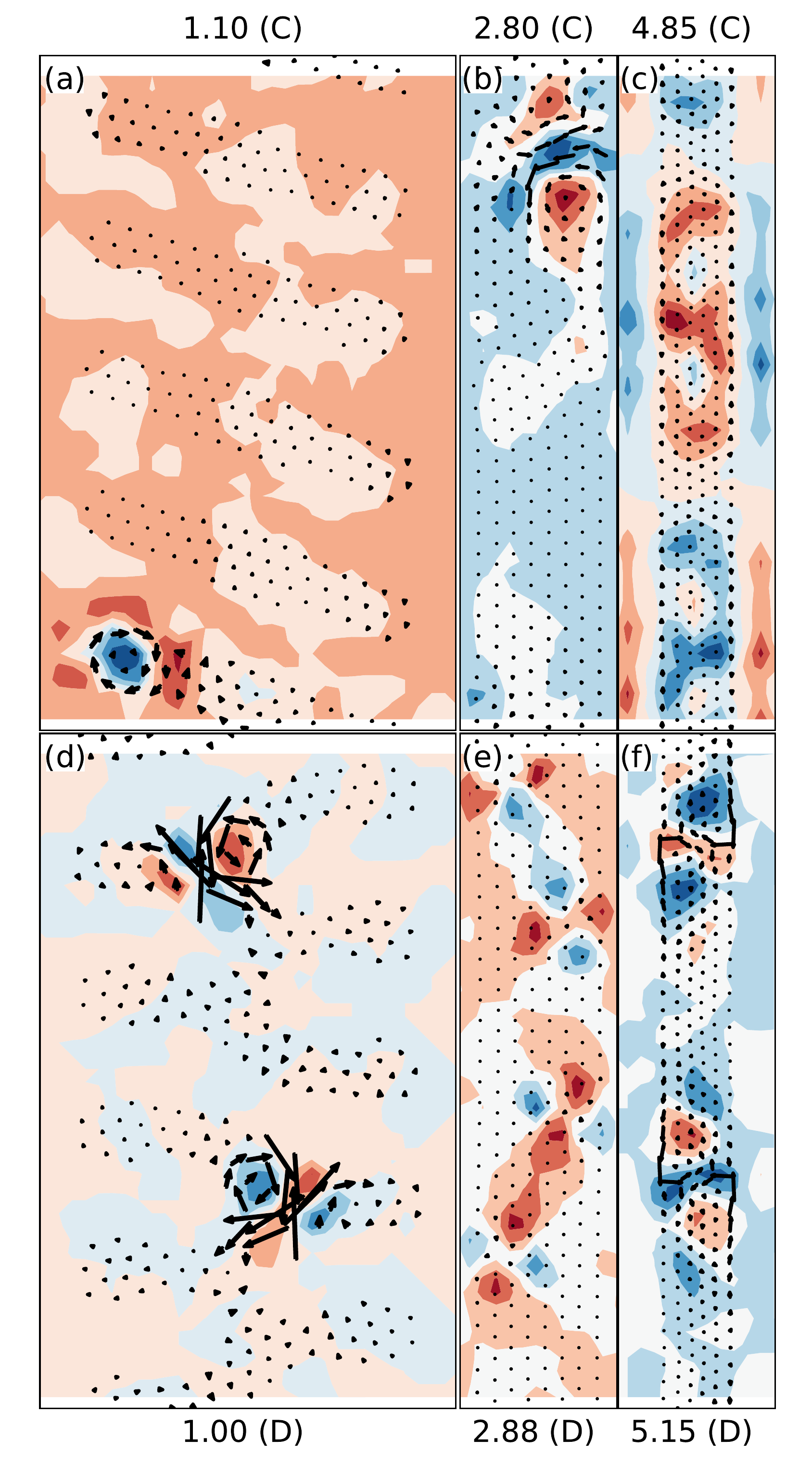}
\caption{The curl $\vec{\omega}$ illustrating the relative symmetry or asymmetry of quadrupole formation under (a-c) compression (C) and (d-f) decompression (D) for increasing $F_p$.  Blue indicates clockwise rotation, red indicates counterclockwise rotation, and the color scale differs from panel to panel. (a) $F_p=1.0$. (b) $F_p=2.8$. (c) $F_p=4.85$. (d) $F_p=1.0$. (e) $F_p=2.88$. (f) $F_p=5.15$.  
}
\label{fig:9}
\end{figure}

Although the range of scaling for the velocity distributions is
too small to unambiguously determine whether a power law tail is present,
we note that studies of the
velocity distribution tails
for 2D particle-based models of dislocation motion
give power laws with exponents
of $\tau = -2.0$ to $-2.5$,
where it is argued that
exponents greater than $\tau=-2.0$
indicate that collective events are important \cite{New1}.
Additionally, 2D simulations of particle-based models of driven dislocations
undergoing avalanches also produce
velocity distributions that can be fit to a power law with
$\tau = -2.5$ \cite{New2}.
In our work, we measure the
velocity of the individual particles,
and not that of dislocations in the packing;
however, in the particle-based models of dislocation motion,
the pairwise dislocation-dislocation interactions
include competing attractive and repulsive terms.

In Fig.~\ref{fig:7}(a-c) we show the 
formation of a quadrupole-like defect 
in the void system under compression 
at $F_p=2.28$.
In addition to the particle velocities, 
illustrated in Fig.~\ref{fig:7}(a), 
we analyze the particle motion   
by mapping it onto a velocity field.  
This allows us to estimate 
the spatial distribution 
of vorticity using 
the curl $\vec{\omega}=\nabla \times \vec{v}$, 
shown in Fig.~\ref{fig:7}(b), 
and the
enstropy, 
or rotational kinetic energy,  
$\epsilon(\omega) = \frac{1}{2} \omega^2$, 
shown in Fig.~\ref{fig:7}(c). 
We overlay $\vec{\omega}$ and $\epsilon$ 
with the particle velocity vectors to show the correlation 
between the field vorticity and the instantaneous particle motion.
From $\vec{\omega}$ in Fig.~\ref{fig:7}(b) we observe that
cooperative particle motion  produces
a quadrupole-like defect.
The compressive substrate force
pushes a portion of the particles
inward near the edge of the open void, 
decreasing the void size 
as shown in Fig.~\ref{fig:1}.  
The short-range attractive force
causes the immediate neighbors of the moving
particles to also move in their wake, whereas
the long-range repulsive force 
pushes more distant particles away.  
Thus the competing interactions 
create a quadrupole-like
defect that does not appear
in similar simulations 
of purely repulsive particles\cite{38}.
These quadrupoles take the form of
two combined defects, 
and have been previously observed 
in simulations of sheared particles 
where the defects were primarily located at 
the edge of the sample subjected to the largest applied shear ~\cite{40}.
Our simulations suggest that 
even in the absence of shear,
quadrupoles tend to form at the edges of the particle assembly
where the local particle density is reduced.
Typically, only a single 
quadrupole-like defect 
appears in the compressed system, 
not only in the void regime
as shown in Fig.~\ref{fig:7}(a-c), but also
in the stripe and dense solid regimes.

In Fig.~\ref{fig:7}(d-f) we show the 
formation of the void state at $F_p=2.24$
under decompression.
The bulk outward flow of the particles suppresses
plastic rearrangement
and prevents the formation of isolated quadrupoles.
Instead, the quadrupoles form in pairs, as shown
in Fig.~\ref{fig:7}(e).
In Fig.~\ref{fig:7}(d), the particle locations and velocity vectors 
indicate that the most rapid motion 
occurs at the edges of the newly forming voids,
as shown by the green particles at the sample center 
and their light blue neighbors.
Above and below this expansion front  
are particles that appear motionless, 
as indicated by their dark blue color. 
Alternating with the rapidly moving regions, and separated from
them by the motionless particles, are groups of
light blue particles that have a small inward velocity.  
In  Fig.~\ref{fig:7}(e) 
we plot the curl of the velocity field 
and 
in Fig.~\ref{fig:7}(f) 
we show the corresponding enstropy 
using the same colors described in Fig.~\ref{fig:7}(a-c), 
but on a different scale. 
Under expansion, the quadrupole defects tend to form in symmetric pairs with
opposite polarity.
This occurs because the decompression is a more orderly process
than compression, so all of the interior regions of the packing can reach
the threshold for quadrupole formation simultaneously, rather than
having just one location reach the quadrupole nucleation threshold prior to the
rest of the sample due to the plastic flow that occurs under compression.
The attractive interaction 
mediates the counter rotation at local length scales, 
while the long range repulsion 
produces cooperative motion across the sample,
giving an ordered pattern of quadrupoles under decompression.
The velocity vectors show that
the inward velocities are quite small under expansion,
producing only weak large scale vorticity
in contrast to the strongly localized motion that generates
strong local vorticity during compression.
A similar distinction between 
large scale rotation 
and localized quadrupole defects was noted in the sheared 
system of Ref.~\cite{40}.

In Fig.~\ref{fig:8} we plot the 
velocity distributions $P(V_x)$
over a force range of $\Delta F_p = 0.03$ 
to highlight the asymmetry 
that appears in $V_x$ 
when a single quadrupole defect forms.
Figure~\ref{fig:8}(a) shows 
the void system under compression 
at $F_p=2.28$,
as illustrated in Fig.~\ref{fig:7}(a-c),  
and
under expansion
at $F_p=2.24$,
as illustrated in Fig.~\ref{fig:7}(d-f).
The magnitude of $V_x$ is much higher 
under compression, 
spanning the range $-0.007 < V_x < 0.014$, 
while the velocities in the expanding system 
fall within the range $-0.002 < V_x < 0.002$.  
Under compression,
$P(V_x)$ 
is heavily skewed toward positive $V_x$ 
since the quadrupole defect 
forms at the right side of the void 
and triggers a net inward motion of particles.
Similarly $P(V_y)$ (not shown) is 
skewed toward positive $V_y$ since the quadrupole forms 
below the void.
Under decompression,
$P(V_x)$ is symmetric, 
consistent with the velocity vectors in 
Fig.~\ref{fig:7}(d).
In Fig.~\ref{fig:8}(b) 
we plot $P(V_x)$ under compression
in the stripe phase 
for $F_p = 1.1$,
as illustrated in Fig.~\ref{fig:9}(a), 
and under decompression for
$F_p = 1.0$,
as illustrated Fig.~\ref{fig:9}(d).
Under compression 
$P(V_x)$ is nearly symmetric,
with a slight skew toward negative $V_x$ 
due to the vortexlike motion 
on the lower left side of the sample
in Fig.~\ref{fig:9}(a).
$P(V_x)$ is also nearly
symmetric under decompression,
with a slight skew to 
positive $V_x$ values.  
These slight asymmetries 
are weak compared to those in the void state
shown in Fig.~\ref{fig:8}(a), 
and indicate that it is possible to identify the presence of
isolated quadrupole defects 
through the asymmetry the generate in $P(V_x)$.

The stripe state under compression at $F_p=1.10$, shown in
Fig.~\ref{fig:9}(a), 
contains a single strong vortex
as described in Fig.~\ref{fig:5}(a).
In contrast,
the labyrinth state under decompression at $F_p=1.0$ exhibits
plastic motion that produces a pair of quadrupoles, as
shown in Fig.~\ref{fig:9}(d). 
There is a large vortex with quadrupole-like features in
Fig.~\ref{fig:9}(b)
under compression
at $F_p=2.8$ where the sample 
transitions into an $n=8$ crystalline triangular lattice. 
The defect forms asymmetrically on one side of the sample
due to the plastic rearrangements.
In contrast, 
Fig.~\ref{fig:9}(e) 
shows that under decompression
at $F_p=2.88$, pairs of weak quadrupoles
form
in a nearly crystalline $n=8$
state.
Figure~\ref{fig:9}(c) illustrates
the large scale motion that occurs under compression
at $F_p=4.85$  
when the system transitions
to a nearly crystalline $n=6$
state,
which is associated with a large negative spike in $dE/dF_p$
and collective particle motion.
Under decompression at $F_p=5.15$,
Fig.~\ref{fig:9}(f) 
shows that a pair of quadrupole-like excitations form.
Such excitations are
common at high confinements, and their formation is enhanced by the
finite width of the particle packing.

\begin{figure}
\includegraphics[width=0.5\textwidth]{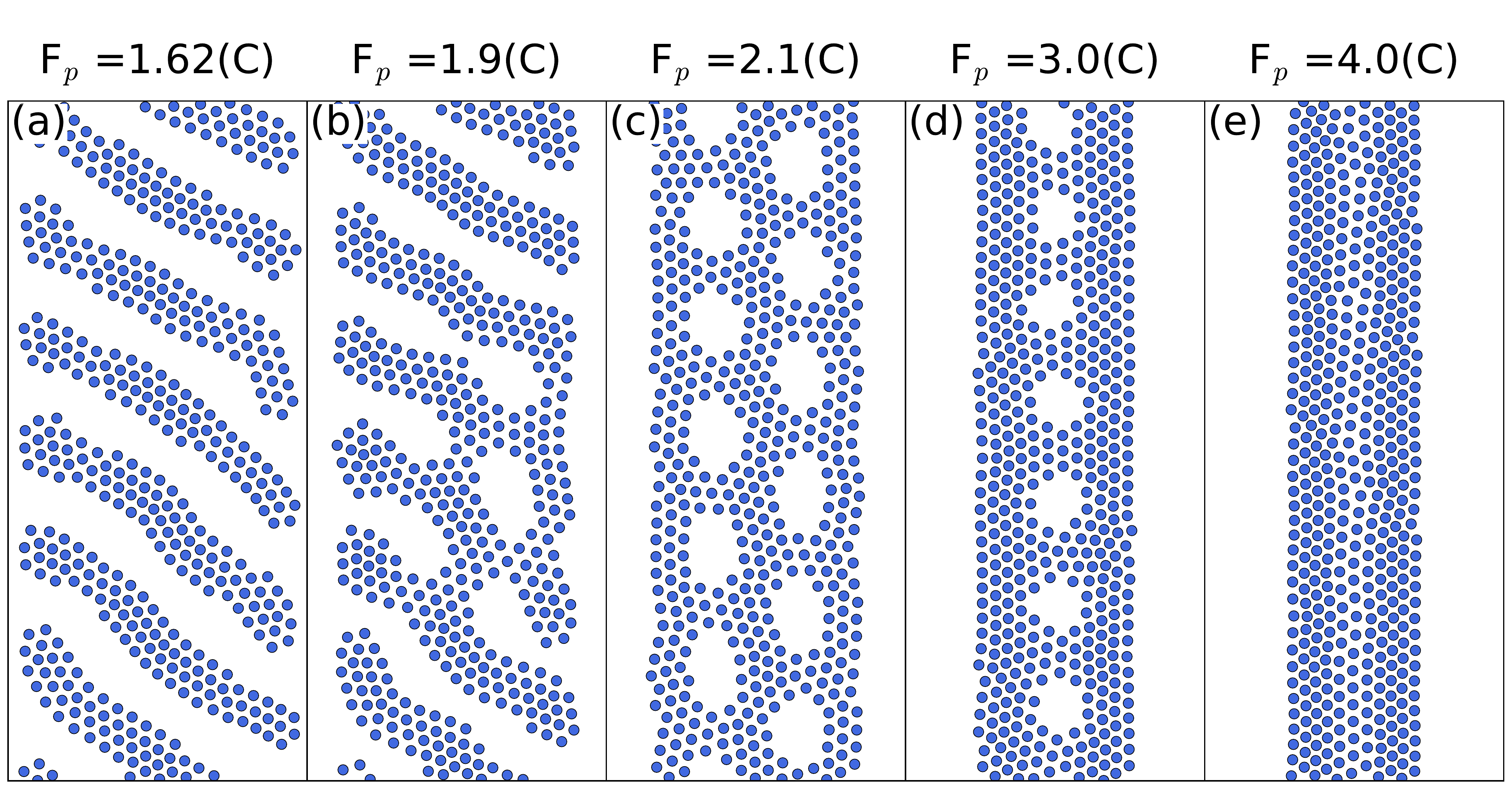}
\caption{Particle positions under compression in a larger system with $L=50$ containing $N=475$ particles. (a) A labyrinth stripe state at $F_p=1.62$. (b) The transition between stripes and voids at $F_p=1.9$. (c) A double row of voids at $F_p=2.1$. (d) A single row of voids at $F_p=3.0$. (e) A close-packed triangular solid at $F_p=4.0$. The supplementary materials fully illustrate the dynamics of compression \cite{supp3}. 
}
\label{fig:10}
\end{figure}

In addition to the $L=36.5$ sample containing $N=256$ particles
considered above, we also
we annealed and compressed $N=475$ particles in a sample
of size $L=50.0$, using the same compression rate described above.
In general, we find that the larger system is more disordered 
due to the diminished 
influence of the short range attractive forces across the system.
In Fig.~\ref{fig:10} we show the particle positions
under compression
in the $L=50.0$ sample
At $F_p=1.62$ in
Fig.~\ref{fig:10}(a), 
the stripe phase is less ordered, and it
persists to higher $F_p$ compared to the smaller sample since greater
compression is required to bring the larger number of particles to a high
enough local density to form voids.
At $F_p=1.9$ in 
Fig.~\ref{fig:10}(b),
the transition to the void state has begun, and we find two instead of one
rows of voids.
We expect that in
larger systems would stabilize larger numbers of rows
of voids,
since annealed systems containing no substrate
form a complete crystal of voids \cite{4}. 
The fully formed double void state appears in 
Fig.~\ref{fig:10}(c) 
at $F_p=2.1$.
The voids are oblong 
and lack a uniform shape and size.
The particle structure 
is also disordered, 
resembling conjoined stripes 
more than a uniform lattice.  
As $F_p$ increases, the local ordering of the particles increases,
but the voids never become uniform in shape.
At $F_p=2.6$, a transition to a single 
row of nonuniform voids occurs,
as shown in Fig.~\ref{fig:10}(d) for $F_p=3.0$. 
The particles surrounding 
the voids form a defected 
triangular lattice.
Like the stripe state, the void state persists to higher $F_p$ in
the larger system, and does not
collapse into a dense triangular lattice
until $F_p=3.6$,
whereas the voids collapse at $F_p=2.4$ in the smaller system.
We show the close packed triangular phase 
at $F_p=4.0$ in Fig.~\ref{fig:10}(e), where there is 
still a reduced density 
at the center of the system which disappears for higher $F_p$.

\section{Summary}

We have studied the compression and decompression in a time-dependent
potential of
a pattern-forming system of particles
that interact via
a short-range attractive force
and a long-range repulsive force.
As a function of increasing confinement, the
system undergoes a series of dynamic rearrangements
from clumps to stripes to voids, and finally forms a dense solid.
The rearrangements occur via slow elastic
motion interspersed with avalanche events of two types.
Large avalanches are associated with structural changes  from
one type of pattern to another, while smaller avalanches occur during
local structural rearrangements
that consist of the motion of
dislocations in the patterns.
The velocity distributions during
large avalanches exhibit strongly non-Gaussian features, and
the tails of the distributions can be fit to
power laws.
At the highest compressions,
the system forms a dense  solid which becomes denser under further compression
through a series of row reduction transitions.
We compare the dynamically produced structures
with those obtained by performing simulated annealing
in a static confining potential of the same strength,
and find that the annealed patterns are more ordered, indicating that
the dynamical  compression produces persistent metastable states.

When we decompress the fully compressed sample, we observe hysteresis
in which
the void, stripe, and clump states reappear at lower confinements
for decompression than those at which they disappeared during compression.
The morphology of some of the patterns is
different during decompression due to the
relative symmetry of tensile forces compared to the asymmetry of the
compressive forces.
We also find that the system undergoes fewer plastic rearrangements upon
decompression than during compression.
The particle velocity distributions
indicate that higher speeds occur more frequently under compression,
so the row reduction avalanches under compression are stronger
than the row expansion avalanches that occur during decompression.
We also observe a difference in the
defects created under compression and decompression,
with
single quadrupole defects forming at the sample edge
during compression and
symmetric pairs of defects appearing in the center of the sample
during decompression.
Our work shows that it may be possible to
control the morphology of the rich phases
of a pattern forming particle system
by using a confining trap to tune the system density.
Such a procedure may allow the creation of
metastable persistent phases that cannot be accessed
in an equilibrium system.

\section{Acknowledgments}
This work was carried out under the auspices of the 
NNSA of the 
U.S. DoE
at 
LANL
under Contract No.
DE-AC52-06NA25396.
The work of DM was supported in part by the U.S. DoE, Office of Science,
Office of Workforce Development for Teachers and Scientists (WDTS) under the
Visiting Faculty Program (VFP).

\footnotesize{
 
}

\begin{thebibliography}{99}
\bibitem{supp1}
  Movie of the compression sequence 
  are available at xx.
\bibitem{supp2}
  Movie of the decompression sequence 
  are available at xx.
\bibitem{supp3}
  Movie of the compression sequence of N = 475
  with particles colored as in Fig.~\ref{fig:5}  
  are available at xx.
\end{thebibliography}
\end{document}